\documentclass[11pt,authorversion=true,sigconf,twocolumn=false]{acmart} 
\usepackage{amsmath, amssymb}
\usepackage{paralist}
\usepackage{listings}
\usepackage{lstlinebgrd}
\lstdefinestyle{pseudocode_style}{
  escapeinside={@|}{|@},
  language=Python,
  basicstyle=\footnotesize\ttfamily,
  showspaces=false,
  commentstyle=\color{green!40!black},
}
\usepackage{balance}
\newcommand{\op}{\ensuremath{\mathsf{op}}}
\newcommand{\inv}{\ensuremath{\mathsf{Inv}}}

\newcommand{\precond}[1][\op{}]{\ensuremath{\mathsf{Pre}_{#1}}}

\newcommand{\mergef}{\ensuremath{\mathsf{\mathtt{merge}}}}

\usepackage{xcolor}

  \title{Invariant Safety for Distributed Applications}
  \author{Sreeja S Nair}
  \affiliation{Sorbonne Universit\'e---LIP6 \& Inria, Paris, France}
  \email{sreeja.nair@lip6.fr}
  \author{Gustavo Petri}
  \affiliation{ARM Research, Cambridge, UK}
  \email{gustavo.petri@arm.com}
  \author{Marc Shapiro}
  \affiliation{Sorbonne Universit\'e---LIP6 \& Inria, Paris, France}
  \email{marc.shapiro@acm.org}
  
  \acmISBN{978-1-4503-6276-4}
  \acmConference[PaPoC'19]{Principles and Practice of Consistency for Distributed Data}{March 25}{Dresden, Germany}
\begin{abstract}
  We study a proof methodology for verifying the safety of data invariants of highly-available distributed applications
  that replicate state.  
  The proof is 
  \begin{inparaenum}
    \item modular: one can reason about each individual operation separately, and 
    \item sequential: one can reason about a distributed application as if it were sequential. 
  \end{inparaenum}
  We automate the methodology and illustrate the use of the tool with a representative example. 
\end{abstract}
        
\begin{document}

\maketitle

\section{Introduction}
\label{sec:intro}
A distributed application often replicates its data to several
locations, and accesses the closest available
replica.
Examples include social networks, multi-user games,
co-operative engineering tools, collaborative editors, source control
repositories, or distributed file systems.
To ensure availability, an update must not synchronise across replicas;
otherwise, when a network partition occurs, the system will block.
Asynchronous updates may cause replicas to diverge or to violate the
data invariants of the application.

To address the first problem, Conflict-free Replicated Data Types
(CRDTs)\cite{syn:rep:sh143} have mathematical properties to ensure that
all replicas that have received the same set of updates converge to the
same state \cite{syn:rep:sh143}.
To ensure availability, a CRDT replica executes both queries and updates
locally and immediately, without remote synchronisation.
It propagates its updates to the other replicas asynchronously.

There are two basic approaches to update propagation: to propagate
operations, or to propagate states.
In the former approach, an update is first applied to some origin replica,
then sent as an operation to remote replicas, which in turn apply it to
update their local state.
Operation-based CRDTs require the the message delivery layer to deliver
messages in causal order, exactly once; the set of replicas must be
known.

In the latter approach, an update is applied to some origin replica.
Occasionally, one replica sends its full state to some other replica,
which merges the received state into its own.
In turn, this replica will later send its own state to yet another
replica.
As long as every update eventually reaches every replica transitively,
messages may be dropped, re-ordered or duplicated, and the set of
replicas may be unknown.
Replicas are guaranteed to converge if the set of states, as a result of
updates and merge, forms a monotonic semi-lattice \cite{syn:rep:sh143}.
Due to these relaxed requirements, state-based CRDTs have better
adoption \cite{BaqueroACF17}.
They are the focus of this work.

As a running example, consider a simple auction system.
The state of an auction consists of status, a set of bids,
and a winner.
This state is replicated at multiple servers; CRDTs ensures that all
replicas eventually converge.
Users at different locations can start an auction, place bids, close the
auction, declare a winner, inspect the local replica, and observe if a
winner is declared and who it is.
All replicas will eventually agree on the same auction status, same set
of bids and the same winner.

However, the application may also require to maintain a correctness
property or \emph{invariant} over the data.
An invariant is an assertion on application data that must evaluate to
true in every state of every replica.
For instance, the auction's invariant is that: when the auction is
closed, there is a winner; there is a single winner; and the winner's
bid is the highest.

Such an invariant is easy to ensure in a sequential system, but
concurrent updates might violate it.
In this case, the application would need to synchronise some updates
between replicas, in order to maintain the invariant.
For instance, in the absence of sufficient synchronisation, a replica
might close the auction and declare a winner, while concurrently a user
at a different replica is placing a higher bid.

This problem has been addressed before, by stating correctness rules and
proof obligations; however, previous work considers only the
operation-based approach \cite{syn:app:sh179,app:sh183,syn:lan:1778}.

In this paper, we propose a proof methodology for applications that use
state-based CRDTs.
We exploit the properties of state-based CRDTs to reason about a
concurrent system in a sequential manner.
We have also developed a tool named Soteria, to automate our proof rule.
Soteria detects concurrency bugs and provides counterexamples.

\section{System Model}
\label{sec:model}
An application consists of state, some operations, a merge function, and
an invariant.
The state is replicated at any number of replicas.
A client chooses any arbitrary replica for its next operation, called
the \emph{origin replica} for that operation.
A replica occasionally sends its state to some other replica, which the
receiving replica \emph{merges} into its own state.
In summary, the state of any given replica changes, either by executing
an update operation for which it is the origin, or by {merging} the
state received from a remote replica.
Each replica is sequential.
A merge is the only point where a replica observes concurrent
operations submitted to other replicas.
Each replica executes sequentially, one local update or merge at a time;
equivalently, an update or merge operation executes atomically, even if
it updates multiple data items.

An application invariant is an assertion over state.
The invariant must evaluate true in every state of every replica.
Despite being evaluated against local state, an invariant is in effect
global, since it must be true at all replicas, and replicas eventually
converge.

\begin{figure*}
  \centering
  \includegraphics[width=0.65\textwidth]{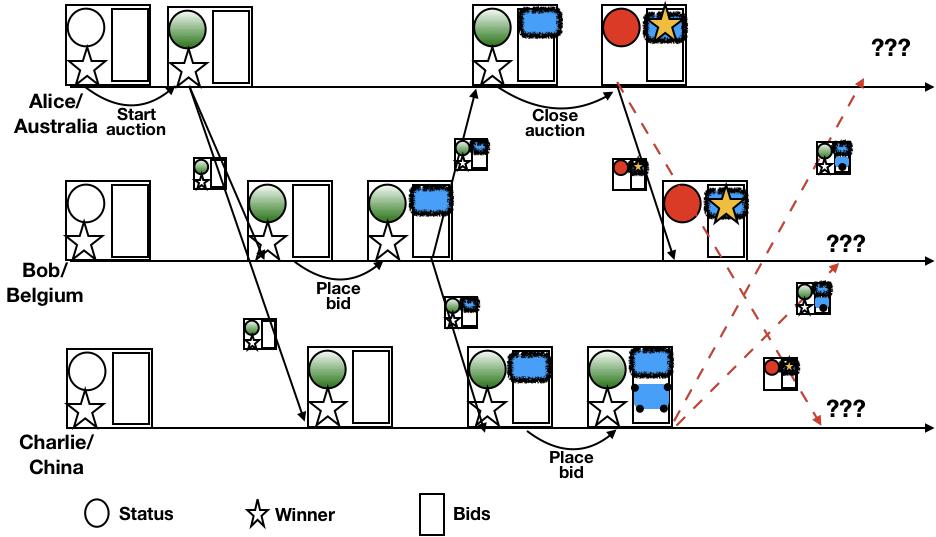}  
  \caption{Evolution of state in an auction application}
  \label{fig:auctionex}
\end{figure*}

\autoref{fig:auctionex} depicts the evolution of state in our auction
application.
Each line represents a replica, time progressing from left to right.
A box represents local state.
A curved arrow represents an update operation, labelled with the
operation name.
A diagonal arrow shows propagation (labelled with the propagated state),
merged at the receiving replica.

We assume here that application state is a composition of CRDTs.
This is not a limitation, since many basic CRDT types have been
proposed, which extend familiar sequential data types with a concurrency
semantics \cite{rep:syn:sh138}.

The CRDT convergence rules, or \emph{lattice rules}, are the following.
The state of a replica progresses monotonically with time.
The set of states forms a semi-lattice, i.e., is equipped with a partial
order and a least-upper-bound function.
A state transition represents the execution of either an update
operation or merge.
An update is an inflation, i.e., the resulting state is no less (in the
partial order) than the previous one.
\emph{Merge} computes a state which is the least-upper-bound of the
current local state and the received remote state.

We write $\sigma$ for the current state of a replica, $\sigma_{new}$ for
the new state after an operation or merge, and $\sigma'$ for the
incoming state from a remote replica.
In the Boogie specification language (used by our tool, described in
Section~\ref{sec:tool}), we denote operation execution with the keyword
\texttt{call}, an assertion with \texttt{assert}, and assumptions with
\texttt{assume}.
An operation $\op$ executes only if its \texttt{precond}ition is true.
Thus, we can write the above lattice conditions as follows:
\begin{itemize}
  \item An update  operation $\op$ is an inflation.
  \begin{lstlisting}[style=pseudocode_style]
call @|$\sigma_{new} = \op(\sigma)$|@   
assert @|$\sigma_{new} \geq \sigma$|@
  \end{lstlisting}
  \item Merge is a least upper bound
  \begin{lstlisting}[style=pseudocode_style]
call @|$\sigma_{new} = \mergef(\sigma, \sigma')$|@  
assert @|$\sigma_{new} \geq \sigma \wedge \sigma_{new} \geq \sigma'$|@  #upper bound
assert @|$\forall \sigma^*, \sigma^* \geq \sigma \wedge \sigma^* \geq \sigma' \implies \sigma^* \geq \sigma_{new}$|@ #least
  \end{lstlisting}
\end{itemize}

Let us illustrate with the auction example ({for simplicity we consider
  a single auction}).
Its state is as follows:
\begin{itemize}
  \item \texttt{Status}: the status of an auction can move from its
    initial state, \texttt{INVALID} (under preparation), to
    \texttt{ACTIVE} (can receive bids) and \texttt{CLOSED} (no more bids
    accepted), such that \texttt{INVALID} $<$ \texttt{ACTIVE} $<$
    \texttt{CLOSED}.
  \item \texttt{Winner}: The winner of the auction.
    It is either $\bot$, initially, or the bid with the highest amount.
    In case of multiple bids with same amount, the bid with the lowest
    id wins.
    It is ordered such that $\forall b \in Bids, \bot \leq b$.
  \item \texttt{Bids}: Set of bids placed (initially empty)
  \begin{itemize}
    \item \texttt{BidId}: A unique identifier for each bid placed
    \item \texttt{Placed}: A boolean flag to indicate whether the bid
      has been placed or not, ordered \texttt{TRUE} $>$ \texttt{FALSE}.
      It is enabled once when the bid is placed.
      Once placed, a bid cannot be withdrawn.
    \item \texttt{Amount}: An integer representing the amount of the
      bid; this cannot be modified once the bid is created.
  \end{itemize}
\end{itemize}

\autoref{fig:auctionex} illustrates how the auction state evolves over
time; status, bids and winner are represented by a circle, rectangle and
a star respectively.
The application state is geo-replicated at data centres in
Australia, Belgium and China.

We now specify the \texttt{merge} operation for an auction.
We denote the receiving replica state $\sigma$ = (status, winner, Bids);
the received state is denoted $\sigma'$ = (status$'$, winner$'$,
Bids$'$).
\begin{lstlisting}[style=pseudocode_style]
  merge((status,winner,Bids),
        (status@|$'$|@,winner@|$'$|@,Bids@|$'$|@)):
    status@|$_{new}$|@ := max(status,status@|$'$|@)
    if winner@|$'$|@@|$\neq$|@@|$\bot$|@ then winner@|$_{new}$|@ := winner@|$'$|@ 
        else winner@|$_{new}$|@ := winner
    @|$\forall$|@b, if b @|$\in$|@ Bids @|$\cap$|@ Bids@|$'$|@ then 
        Bids@|$_{new}$|@.b.placed := Bids.b.placed 
                        @|$\vee$|@ Bids@|$'$|@.b.placed
      else if b @|$\in$|@ Bids@|$'$|@ then 
        Bids@|$_{new}$|@.b.placed := Bids@|$'$|@.b.placed
      else Bids@|$_{new}$|@.b.placed := Bids.b.placed
    @|$\forall$|@b, if b @|$\in$|@ Bids then 
        Bids@|$_{new}$|@.b.amount := Bids.b.amount
      else Bids@|$_{new}$|@.b.amount := Bids@|$'$|@.b.amount
\end{lstlisting}

In the absence of any extra synchronisation, it is possible to violate
the invariant, as the following execution scenario illustrates.
Alice from Australia starts an auction, by setting its status  to
\texttt{ACTIVE} (green in the figure) in the Australian replica.
Henceforth, the auction can receive bids.
The Australian replica sends its updated state to the other two replicas,
which merge it into their own states.
Now Bob in Belgium places a bid for \$100 (blue).
This update is sent to other replicas.
Charlie, in China, sees the auction and Bob's bid.
He updates the China state with a higher bid of \$105 (dotted blue),
which is sent to both other replicas.
However, due to a network failure, the remote replicas do not receive
this update (red dotted lines).
Meanwhile Alice, unaware of Charlie's bid, closes the auction and
declares Bob's bid as the winner.
Later when the network heals, the updated states are sent and
merged.
The auction is new closed, and contains Bob's \$100 bid and Charlie's
\$105 bid.
Unfortunately, Bob's bid is the winner, violating the application
invariant.

In the next section, we can discuss how to ensure invariants of
applications build on top of the system model we described.

\section{Proving Invariants}
\label{sec:rule}

As explained earlier, each replica executes a sequence of state
transitions, due either to a local update, or to a merge incorporating
remote updates.
Thus, concurrency can be observed only through merge.

Let us call \emph{safe state} a replica state that satisfies the
invariant.
Assuming the current state is safe, any update (local or merge) must
result in a safe state.
To ensure this, every update is equipped with a precondition that
disallows any unsafe execution.%
\footnote{
  Technically, this is at least the weakest-precondition of the update
  for the invariant.
  It strengthens any \emph{a priori} precondition that the developer may
  have set.
}
Thus, a local update executes only when, at the origin replica, the
current state is safe and its precondition currently holds.
Similarly, merge executes only with two safe states that together
satisfy a merge precondition.

Formally, an update $u$ (an operation or a merge), mutates the local
state $\sigma$, to a new state $\sigma_{new} = u(\sigma)$.
To preserve the invariant, \texttt{Inv}, we require that
\[\sigma \in \precond[u] \implies u(\sigma) \in Inv\]

To illustrate local preconditions, consider an operation
\texttt{close\_auction(w: BidId)}, which sets auction status to
\texttt{CLOSED} and the winner to \texttt{w}.
The developer may have written a precondition such as $\texttt{status} =
\texttt{ACTIVE}$, because closing an auction doesn't make sense otherwise.
In order to ensure the invariant that the winner has the highest amount,
one needs to strengthen it with the clause
$\texttt{is\_highest(Bids, w)}$, defined as 
$\forall \texttt{b} \in
\texttt{Bids}: \texttt{b.Amount} \le \texttt{w.Amount}$.

\newcommand{\incn}{\ensuremath{\texttt{incn}}}
\newcommand{\incm}{\ensuremath{\texttt{incm}}}
\newcommand{\merge}{\ensuremath{\texttt{merge}}}
To illustrate merge precondition, consider a CRDT whose state is
the pair of integers, $\sigma = (n,m) \in \mathbb{N} \times \mathbb{N}$.
It has two operations, \incn{} and \incm{}, that respectively increment
$n$ or $m$ by 1, and a merge function:
\[\merge(\sigma,\sigma') = (\max(n,n'), \max(m,m'))\]
We wish to maintain the invariant that their sum is no more than 10:
\[Inv \triangleq (n+m) \le 10\]
The precondition of \incn{} is $\precond[\incn] \triangleq (n+m) \le 9$;
similarly for \incm.
Starting from a safe state $(4, 5)$, two replicas may independently
increment to states $(5,5)$ and $(4,6)$ respectively.
Both are safe.
However, merging them would violate the invariant.
Therefore, $\merge(\sigma,\sigma')$ must have precondition
\[\precond[\mergef] \triangleq \max(n,n') + \max(m,m') \le 10\]

Since merge can happen at any time, it must be the case that its
precondition is always true, i.e., it constitutes an additional
invariant.
Now our global invariant consists of two parts: first, the application
invariant, and second, the precondition of merge.
We can now state our proof rule informally as follows::%
\footnote{
  We omit the full formalisation and the proof of soundness for brevity.
}
\begin{proof}[State-Based Safety Rule]\bf
  Define the \emph{precondition of merge} to be the weakest-precondition
  of merge, for the application invariant.
  The initial state must satisfy, and each local update or merge
  operation must preserve, the conjunction of:
    \begin{inparaenum}[\em (i)]
    \item
      the application invariant, and~
    \item
      the precondition of merge.
    \end{inparaenum}
\end{proof}
In our Boogie notation, each operation can be verified as follows:
\begin{lstlisting}[style=pseudocode_style]
  assume @|$\inv \wedge \precond[\mergef] \wedge \precond$|@
  call @|$\sigma_{new} = \op(\sigma)$|@
  assert @|$\inv \wedge \precond[\mergef]$|@
\end{lstlisting}
The case of the $\mergef$ function can be verified with the
following condition:
\begin{lstlisting}[style=pseudocode_style]
  assume @|$\inv \wedge \inv' \wedge \precond[\mergef]$|@
  call @|$\sigma_{new} = \mergef(\sigma, \sigma')$|@
  assert @|$\inv \wedge \precond[\mergef]$|@
\end{lstlisting}
Note that there are two copies of state, the unprimed local state of the
replica applying the merge, and the
primed state received from a remote replica.
Inv$'$ denotes that $\sigma'$ preserves the invariant Inv.

\subsection{Applying the proof rule}

Let us apply the proof methodology to the auction application.
Its invariant is the following conjunction:
\begin{enumerate}
\item \label{item:active}
  A bid is placed only when status is \texttt{ACTIVE}.
\item  \label{item:nochange}
  And: Once a bid is placed, its amount does not change.
\item \label{item:closed}
  And: There is no winner until status is \texttt{CLOSED}.
\item  \label{item:highest}
  And: There is a single winner, the bid with the highest amount (breaking
  ties using the lowest identifier).
\end{enumerate}
Computing the weakest-precondition of each update operation, for this
invariant, is obvious.
For instance, as discussed earlier, \texttt{close\_auction(w: BidId)}
gets precondition $\texttt{is\_highest(Bids, w)}$, because of
Invariant Term~\ref{item:highest} above.

Despite local updates to each replica preserving the invariant,
\autoref{fig:auctionex} showed that it is susceptible of being violated
by merging.
This is the case if Bob's \$100 bid in Belgium wins, even though Charlie
concurrently placed a \$105 bid in China; this occurred because 
\texttt{status} became \texttt{CLOSED} in Belgium while still
\texttt{ACTIVE} in China.
The weakest-precondition of merge for Term~\ref{item:highest} expresses
that, if \texttt{status} in either states is \texttt{CLOSED}, the winner
should be the bid with the highest amount in both the states.
Therefore, $\merge(\sigma,\sigma')$ must have the following additional
precondition:
\begin{lstlisting}[style=pseudocode_style]
  status=CLOSED @|$\implies$|@ is_highest(Bids, winner)
                  @|$\wedge$|@ is_highest(Bids@|$'$|@, winner)
  @|$\wedge$|@ status@|$'$|@=CLOSED @|$\implies$|@ is_highest(Bids, winner@|$'$|@)
                    @|$\wedge$|@ is_highest(Bids@|$'$|@, winner@|$'$|@)
\end{lstlisting}

Furthermore, the code for merge uses Term~\ref{item:nochange}, for which
its weakest-precondition is as follows:
\begin{lstlisting}[style=pseudocode_style]
  @|$\forall$|@b @|$\in$|@ Bids@|$\cap$|@Bids@|$'$|@, Bids.b.amount = Bids@|$'$|@.b.amount
\end{lstlisting}

These two merge preconditions now strengthen the global invariant, in
order to preserve safety in concurrent executions.
Let us now consider how this strengthening impacts the local update
operations.
Since starting the auction doesn't modify any bids, this operation
trivially preserves it.
Placing a bid might violate it, if the auction is concurrently
closed in some other replica; conversely, 
closing the auction could violate it, if a
higher bid is concurrently placed in a remote replica.
Thus, the auction application is safe when executed sequentially, but is
unsafe when updates are concurrent.
This indicates the specification has a bug, which we now proceed to fix.

\subsection{Concurrency Control for Invariant Preservation}

As we discussed earlier, the preconditions of operations and merge are
strengthened in order to preserve the invariant.
This provides a sequentially safe specification.
An application must also preserve the precondition of merge in order to
ensure concurrent safety.
Violating this indicates the presence of a bug in the specification.
In that case, the developer needs to strengthen the application by
adding appropriate concurrency control mechanisms, ie., the operations
that fail to preserve the precondition of merge might need to
synchronise.
The required concurrency control mechanisms are added as part of the
state in our model.
The modified application state is now composed of the CRDTs that
represents the state and the concurrency control mechanism.
Together, it behaves like a composition of state-based CRDTs.
The whole state should now ensure the lattice conditions described in
\autoref{sec:model}.

Recall that in the auction example, placing bids and closing the auction
were not preserving the precondition of merge.
This requires strengthening the specification by adding a concurrency
control mechanism to restrict these operations.
We can enforce them to be strictly sequential, thereby avoiding
concurrency at all.
But this will affect the availability of the application.

A concurrency control can be better designed with the workload
characteristics in mind.
For this particular application, we know that placing bids are very
frequent operations than closing an auction.
Hence we try to formulate a concurrency control like a readers-writer
lock.
In order to realise this we distribute tokens to each replica.
As long as a replica has the token, it can allow placing bids.
Closing the auction requires recalling the tokens from all replicas.
This ensures that there are no concurrent bids placed and thus a winner
can be declared, respecting the application safety.

The entire specification of the auction application can be seen in
\autoref{fig:auction}.
The shaded lines in blue indicate the effect of adding concurrency
control to the state.

An alternative approach to our treatment of concurrency control could be
to consider the \emph{invariant as a resource} in the style of
Concurrent Separation Logic \cite{OHearn07}.
In this case, access to the application state, described through a
separation logic invariant, is guarded by a concurrency control
mechanism (typically some form of a lock).
However, this approach is tied to separation logic reasoning, where
assertions act as resources, and allows one to distinguish local from
global resources.
We consider that this was not essential for the kind of proofs that we
conduct, but it might be more promising when verifying client programs
of our data types.

\section{Automating the verification}
\label{sec:tool}

In this section we present a tool to automate the verification of
invariants as discussed in the previous sections.
Our tool, called \emph{Soteria} is based on the Boogie \cite{Barnett:2005:BMR:2090458.2090481} verification framework.
The input to Soteria is a specification of the application written in Boogie, 
an intermediate verification language.

A specification in Soteria will consist of the following parts:
\begin{itemize}
  \item \textbf{State:} a declaration of the state. It can be a single CRDT or a composition of CRDTs.
    \item \textbf{Comparison function:}  The programmer
    provides a comparison function (annotated with keyword \texttt{@gteq})
    that determines the partial order on states.
    \item \textbf{Operations:} The programmer provides the
  implementation of the operations and their respective preconditions, $\precond$.
  Operations are encoded either imperatively as Boogie procedures or declaratively as postconditions.
  \item \textbf{Merge function:} The special $\mergef$
  operation is distinguished from the other operations (with annotation \texttt{@merge}). 
  The programmer must provide a precondition to
  $\mergef$ that is strong enough to prove the invariant. 
  \item \textbf{Application Invariant:} The programmer provides the invariant (with keyword \texttt{@invariant}) to be verified by the tool as a Boogie assertion over the state.
\end{itemize}
  In addition, Boogie often requires additional
  information such as: 
  \begin{inparaitem}
  \item User-defined data types, 
  \item Constants to declare special objects such as the 
  origin replica \lstinline|me|, or to
    bound the quantifiers, 
  \item Axioms for inductive functions
    over aggregate data structures, for instance, 
    to compute the maximum of a set of values, 
  \item Loop invariants.
  \end{inparaitem}

  The specification of the auction application can be seen in \autoref{fig:auction}.
  \begin{figure}[hp]
    \begin{center}
        \label{spec:auction}
            \begin{lstlisting}[style=pseudocode_style, 
                linebackgroundcolor={%
                \ifnum\value{lstnumber}=4
                        \color{blue!20}
                \fi
                \ifnum\value{lstnumber}=9
                    \color{blue!20}
                \fi
                }]
Initial state: 
  status = INVALID @|$\wedge$|@ winner = @|$\bot$|@ 
  @|$\wedge$|@ @|$\nexists$|@ b, Bids.b.placed 
  @|$\wedge$|@ @|$\forall$|@ r, Tokens.r = true

Comparison function:
  status@|$_1$|@ @|$\geq$|@ status@|$_2$|@ @|$\wedge$|@ (winner@|$_1$|@@|$\ne$|@@|$\bot$|@ @|$\vee$|@ winner@|$_2$|@=@|$\bot$|@)
  @|$\wedge  \forall$|@b, (Bids@|$_1$|@.b.placed @|$\vee$|@ @|$\neg$|@Bids@|$_2$|@.b.placed)
  @|$\wedge$|@ (@|$\forall$|@r, @|$\neg$|@Tokens@|$_1$|@.r @|$\vee$|@ Tokens@|$_2$|@.r)
            \end{lstlisting}
        \begin{lstlisting}[style=pseudocode_style, 
            linebackgroundcolor={%
            \ifnum\value{lstnumber}=7
                    \color{blue!20}
            \fi
            \ifnum\value{lstnumber}=15
                    \color{blue!20}
            \fi
            \ifnum\value{lstnumber}=16
                    \color{blue!20}
            \fi
            \ifnum\value{lstnumber}=17
                    \color{blue!20}
            \fi
            \ifnum\value{lstnumber}=18
                    \color{blue!20}
            \fi
            \ifnum\value{lstnumber}=19
                    \color{blue!20}
            \fi
            \ifnum\value{lstnumber}=20
                    \color{blue!20}
            \fi
            \ifnum\value{lstnumber}=21
                    \color{blue!20}
            \fi
            \ifnum\value{lstnumber}=25
                    \color{blue!20}
            \fi
            }
            ]
Invariant:
  @|$\forall$|@b, Bids.b.placed @|$\implies$|@ status@|$\geq$|@ACTIVE 
      @|$\wedge$|@ Bids.b.amount>0
  status@|$\leq$|@ACTIVE @|$\implies$|@ winner=@|$\bot$|@
  status=CLOSED @|$\implies$|@ Bids.winner.placed 
    @|$\wedge$|@ is_highest(Bids, winner)
  status=CLOSED @|$\implies$|@ @|$\forall$|@r,@|$\neg$|@Tokens.r
  
{Pre@|$_\mathtt{merge}$|@:
  status=CLOSED @|$\implies$|@ is_highest(Bids, winner)
            @|$\wedge$|@is_highest(Bids@|$'$|@, winner)
  @|$\wedge$|@ status@|$'$|@=CLOSED @|$\implies$|@ is_highest(Bids, winner@|$'$|@)
            @|$\wedge$|@is_highest(Bids@|$'$|@, winner@|$'$|@)
  @|$\wedge$|@ @|$\forall$|@b, Bids.b.amount = Bids@|$'$|@.b.amount
  @|$\wedge$|@ @|$\forall$|@r, Tokens.r.me @|$\implies$|@ Tokens@|$'$|@.r.me 
  @|$\wedge$|@ @|$\forall$|@r,b, (@|$\neg$|@Tokens.r @|$\wedge$|@ @|$\neg$|@Bids.b.placed) 
              @|$\implies$|@ @|$\neg$|@Bids@|$'$|@.b.placed
  @|$\wedge$|@ @|$\forall$|@ r,b, (r@|$\neq$|@me @|$\wedge$|@ @|$\neg$|@Tokens.r @|$\wedge$|@ @|$\neg$|@Bids.b.placed) 
              @|$\implies$|@ @|$\neg$|@Bids@|$'$|@.b.placed
  @|$\wedge$|@ @|$\forall$|@ r,@|$\neg$|@Tokens.r @|$\implies$|@ winner@|$'$|@=winner @|$\vee$|@ winner@|$'$|@=@|$\bot$|@
  @|$\wedge$|@ @|$\forall$|@ r, Tokens.r @|$\implies$|@ winner=@|$\bot$|@ @|$\wedge$|@ winner@|$'$|@=@|$\bot$|@}
merge((status, winner, Bids, Tokens), 
        (status@|$'$|@,winner@|$'$|@,Bids@|$'$|@,Tokens@|$'$|@)):
  <merge of status, winner, Bids as @|{\footnotesize\texttt{in}}|@ @|\autoref{sec:model}|@>
  @|$\forall$|@r, Tokens@|$_{new}$|@.r := Tokens.r@|$\wedge$|@Tokens@|$'$|@.r
\end{lstlisting}
\begin{lstlisting}[style=pseudocode_style,
    linebackgroundcolor={%
\ifnum\value{lstnumber}=2
        \color{blue!20}
\fi
\ifnum\value{lstnumber}=9
        \color{blue!20}
\fi
\ifnum\value{lstnumber}=15
\color{blue!20}
\fi
}]
{Pre@|$_\mathtt{start\_auction}$|@: status = INVALID @|$\wedge$|@ winner = @|$\bot$|@ 
  @|$\wedge$|@ @|$\forall$|@r, Tokens.r}
start_auction():
  status@|$_{new}$|@ := ACTIVE
  winner@|$_{new}$|@ := @|$\bot$|@
{Pre@|$_\mathtt{place\_bid}$|@: @|$\neg$|@Bids.b_id.placed 
  @|$\wedge$|@ Bids.b_id.amount = value 
  @|$\wedge$|@ status = ACTIVE @|$\wedge$|@ winner = @|$\bot$|@ 
  @|$\wedge$|@ Tokens.me} 
place_bid(b_id, value):
  Bids@|$_{new}$|@.b_id.placed := true
  Bids@|$_{new}$|@.b_id.amount := value
{Pre@|$_\mathtt{close\_auction}$|@: status = ACTIVE @|$\wedge$|@ winner = @|$\bot$|@
  @|$\wedge$|@ Bids.w.placed @|$\wedge$|@ is_highest(Bids, w) 
  @|$\wedge$|@ @|$\forall$|@r, @|$\neg$|@Tokens.r}
close_auction(w):
  status@|$_{new}$|@ := CLOSED
  winner@|$_{new}$|@ := w
\end{lstlisting}
\end{center}
\caption{Specification of auction application}
\label{fig:auction}
\end{figure}

\subsection*{Verification}
The verification of a specification is performed in multiple stages; in order:
\begin{enumerate}
\item \textbf{Syntactic checks}: validates the specification for syntactical errors and checks whether the pre/post conditions are sound.
\item \textbf{Compliance check}: checks whether
  the specification provides all the elements explained earlier.
\item \textbf{Convergence check}: 
checks whether the specification respects the properties of a state-based CRDT, ie.,
each operation inflates the state and merge is the least upper bound.
\item \textbf{Safety check}:
verifies the safety of the application invariant, as
discussed in \autoref{sec:rule}. 
This stage is divided further into two sub-stages:
\begin{itemize}
    \item \textit{Sequential safety:} whether each individual operation (or merge) upholds the invariant.
		If not, the designer
    needs to strengthen the precondition of the corresponding operation (or merge) 

    \item \textit{Concurrent safety:}
    whether every operation (and merge) upholds the precondition of
    merge.
    Note that, while this check relates to the concurrent behaviour of
     state-based CRDTs, the check itself is completely
    sequential, ie., it does not require reasoning about
      operations performed by other processes.
    This check ensures that
    the invariant remains safe during concurrent operation. 
		If this check fails, the application
    needs stronger concurrency control.
\end{itemize}
\end{enumerate}

Each check in Soteria \footnote{The tool along with some sample specifications can be accessed at \url{https://github.com/sreeja/soteria_tool}.}
generates counterexamples when the verification fails. 
These counterexamples might guide the developer in debugging the specification according to the verification steps.

\section{Related Work}
\label{sec:literature}
Several works have concentrated on the formalisation and specification
of eventually consistent
systems \cite{app:rep:1716,syn:1732,formel:rep:1802}.
A number of works concentrate on the specification and correct
implementation of
CRDTs \cite{DBLP:journals/afp/GomesKMB17,DBLP:conf/esop/JagadeesanR18}.
Our work also verifies the CRDT (lattice) conditions, but additionally
verifies an arbitrary application invariant over a replica state.

\citet{syn:app:sh179} provides a proof methodology
for proving invariants of CRDTs that propagate operations.
The associated tools \cite{app:sh183, syn:lan:1778} 
performs the check using an SMT solver as the backend and 
\citet{lan:formel:sh203} discusses some concurrency control suggestions by using the counterexamples generated by the failed proofs.
\citet{syn:app:sh179} assume that the underlying network
ensures causal consistency, and their methodology requires reasoning about concurrent behaviours.
This requires checks for each pair of operations in the application
(reflected as stability verification conditions).
\citet{syn:app:sh179} uses an abstract notion of \emph{tokens} as concurrency control mechanisms. The operations acquire tokens in order to preserve the application invariant.

In contrast, Soteria focuses on state-based CRDTs.
We check convergence by verifying the lattice conditions
of \autoref{sec:model}
 and that because of the rules shown
in \autoref{sec:rule}, we can reduce the problem of
verifying the invariant to sequential proof obligations.
This is reflected by the fact that all of our proofs are standard pre/post
conditions checks using the Boogie framework.
Boogie framework.
In contrast with \citet{syn:app:sh179}, Soteria includes concrete specification of concurrency control as part of the application state.

To the best of our knowledge, ours is the first attempt in automated verification of
invariants of state-based CRDTs. 

\section{Conclusion}
We have presented a proof methodology to verify invariants of
state-based CRDT implementations guaranteeing:
\begin{inparaenum}
\item that the implementation satisfies the lattice conditions of state-based CRDTs
\cite{BaqueroACF17}, and
\item that the implementation satisfies programmer provided invariant 
  reducing the problem to checking that each operation of the data type
  satisfies a precondition of the $\mergef$ function of the state. 
\end{inparaenum}

We implemented Soteria, a tool sitting on top of the Boogie
verification framework, to specify the implementation, its
invariant and validate it.

In future work, we plan to automate concurrency control synthesis.
The synthesised concurrency control can be analysed 
and adjusted dynamically 
to minimise the cost of synchronisation.
Another direction for future work can be to 
decouple the update propagation mechanism of CRDT
from the proof rule resulting in a generic proof rule 
to verify distributed systems.

  \begin{acks}
    The authors would like to thank the anonymous reviewers for their comments which helped in improving this paper.
This research is supported in part 
       by the RainbowFS project
       (\emph{Agence Nationale de la Recherche}, France, number 
       ANR-16-CE25-0013-01) %
   and by European H2020 project
    \href{https://www.lightkone.eu/}{732\,505 LightKone} (2017--2020).%
\end{acks}

\balance
  \bibliographystyle{abbrvnat}

  \bibliography{predef,shapiro-bib-perso,shapiro-bib-ext,references}

\end{document}